\documentclass[conference]{IEEEtran}
\IEEEoverridecommandlockouts
\usepackage{cite}
\usepackage{amsmath,amssymb,amsfonts}
\usepackage{algorithmic}
\usepackage{graphicx}
\usepackage{textcomp}
\usepackage{xcolor}
\def\BibTeX{{\rm B\kern-.05em{\sc i\kern-.025em b}\kern-.08em
    T\kern-.1667em\lower.7ex\hbox{E}\kern-.125emX}}
\begin{document}

\title{Optimists at Heart: \\Why Do We Research Game AI?\\(Extended Version)\\
\thanks{$^1$ This research was conducted while undertaking industrial action as part of Queen Mary University of London's branch of the University and College Union. Part of this action was aimed at the UCU's `four fights' -- Pay, Workload, Equality and Casualisation \cite{fourfights}. These enormous problems faced by our sector change the type of work that we can do as researchers, which this very paper is about. The strike action was also a response to Queen Mary's attempt to deduct 100\% of striking academics' pay even after they returned to work, threatening the livelihoods of their most precariously employed staff \cite{qmstrike22}. I am deeply indebted to my colleages in the QMUCU branch, who were supportive and kind to me during the strike action this year.}
}


\author{\IEEEauthorblockN{Michael Cook}
\IEEEauthorblockA{\textit{Knives and Paintbrushes / Queen Mary University of London$^1$} \\
mike@knivesandpaintbrushes.org
}
}

\maketitle

\begin{abstract}
In this paper we survey the motivations behind contemporary game AI research by analysing individual publications, the researchers themselves, and the institutions that influence them. In doing so, we identify some negative effects on our field, caused both by external forces outside of our control as well as institutionalised behaviours that are easily overlooked. We suggest how we might begin to address some of these issues as a community, and reassert ourselves as the primary driving force behind the field.
\end{abstract}


\section{Introduction}
Game artificial intelligence research stretches back to before the term `artificial intelligence' was in common use \cite{turingChess}, but our specific understanding of the field today, with dedicated workshops, conferences and research groups, is perhaps between twenty and twenty-five years old. The development of the field was put into fast-forward by the recent gold rush in AI which began approximately a decade ago, especially because the early AI boom focused on games such as Go \cite{alphago}, DOTA 2 \cite{openaifive} and the Atari \cite{atari} as key application domains, thus driving interest in games as a platform for AI research.

Research fields need agendas, manifestos and direction. While most fields do not have a single governing body defining this, we can intuit what drives research in a field by examining the events, institutions and individuals that comprise the area. Understanding this drive is vital to understanding the spirit of the field: what it values; what motivates new ideas and directions; and where it is going next. There is also ample evidence to suggest that the wrong kind of incentive can motivate bad quality research \cite{natselection} and harm the careers of junior researchers who are swept up in a wave of metrics, targets and partnerships \cite{chapman}. Understanding how researchers in the area are motivated is also key to understanding how we can change that motivation -- what systems and institutions we can build to direct the field in new directions.

This paper aims to examine, and then explain, some of the motivations behind game AI research. It identifies some of the key factors that drive us both generally as academics, and specifically as game AI researchers. We offer a critique of these motivations, and examine the hidden biases and problems that they harbour, and we identify those who benefit most from our current aims as a community. We also attempt to identify motivations and goals that we may be ignoring, and propose that we reorient ourselves as a field to better achieve them.

Game AI is currently in transition away from being the key battleground for the AI boom \cite{wiggers22}. While we still benefit greatly from the success and proliferation of AI research in general, we are no longer the top domain for major companies to invest in and study. However, we still wield a unique position in AI research, in terms of our relationship with both our partner industry, the general public, and the rest of AI \cite{cook21}. This is a useful moment to reflect on what we want to achieve as a community, and how best we can achieve it. This paper aims to contribute one perspective on this question.


\section{Background -- The State of Game AI} \label{sec:bg}
In this section we set the scene for the paper, describing some of the pressures and situations our work in game AI finds itself today. We avoid discussing the direct motivational consequences of these factors until a later section.

\subsection{The Commercial Games Industry}
The commercial games industry is rapidly changing, and the speed and ferocity of change seems to accelerate with each passing year. Platform holders, publishers and AAA developers are struggling to come to terms with large shifts in the sector's landscape, like increasing unionisation of the workforce \cite{paul22}, pressure on supply of critical technology \cite{williams22}, and the approaching existential threat of climate change \cite{hern22}.

At the same time, being a technology-driven industry, they must maintain constant vigilance and proactivity watching for the next big trend in order to get out ahead of it. Over the last decade we have seen many new trends hailed as the future of games, including virtual reality, cloud gaming, blockchain games, machine learning, eSports and procedural generation. All resulted in huge gambles by major companies, with only a few success stories.

\subsection{Goverment Expansion}
Artificial intelligence is now a major policy area for national governments and international alliances. Spending on AI within both China and the US is under close scrutiny worldwide, with both nations' governments seeking to position themselves as a world leader. The EU is investing in large new AI centres to attract key research talent and retain their work in the space. For better or worse, any of the world's biggest economies see AI as a key technology for many important future issues, including industrial output, defence and security and social care\cite{ukaistrat}.

Despite the general trend of open (or mostly open) access in AI through sites like arXiv, there are clear advantages for governments to retaining AI research within a national area of influence. AI research talent leads to startups and spinoff technology being created within the same country; it can lead to research projects funded to benefit military and security organisations that would be more complex or impossible with overseas partners; and it can compound talent by attracting students and researchers seeking high-profile institutions, and companies wishing to hire employees with specific skills. 

\subsection{Precarity and Boom in Academia}
In 2021 the OECD published a report on precarious workers in academia -- those who have obtained a PhD but not a permanent job \cite{oecd}. It indicates that precarity has greatly increased in OECD nations, and points to several causes for this, including an ageing academic population and higher rates of PhD intake and awards. 75\% of researchers surveyed across the OECD under the age of 35 were on a temporary contract, and researchers were twice as likely as the OECD average to be living in a country other than their country of birth.

%
At the same time, the AI boom presents unique opportunities to those with the right background. Senior AI researchers in particular are frequently poached by both startups and Big Tech, and opportunities for funding startups are greater than ever. The AI boom has also resulted in a surge in applications to study AI, with bespoke courses and institutes being opened, which has resulted in greatly increased pressure on teaching staff, which hiring efforts are not alleviating \cite{ucuworkload}. The changing state of AI is widening the gap between the comfortable senior staff -- who hold permanent positions, and are able to network, consult and take well-paid industry positions -- and the exploited junior staff, who are underpaid, overworked, and must sacrifice more and more to secure a permanent job.

\section{Motivations for Work}
In this section we outline some common motivations given for research projects in the area. We surveyed every paper published at the 2021 IEEE Conference on Games related to AI, and looked for implicit or explicit mentions of motivation for the work. This includes descriptions of existing problems in the field, and potential applications for the work.


\subsection{Cost} \label{sec:cost}
Reduction of cost is a very commonly given type of motivation for game AI research. Papers often frame the problems they are trying to solve as being costly for game developers, sometimes described in terms of hours spent solving a problem, sometimes in terms of the cost in hardware to solve or create something. For example, \cite{lipsynch} explains that traditional approaches to lip-synching in game development are `time-consuming and expensive', thus motivating a need for better approaches to automating this task.

Scientific advances and new technology have a long history of being motivated by a desire to drive down the cost of processes or make them more efficient \cite{marx}, and the familiarity of this reasoning may be a reason why it is popularly used as a motivation in a field dominated by commercial concerns. Despite this, the motivations of cost reduction are not often justified. A classic example of this is procedural content generation, which is frequently motivated as a way to solve the `content generation problem' \cite{contentproblem}. However, we have not seen any game AI paper that justifies the claim that automation saves time or money. Rather, game developers often joke that procedural generation allows game developers to `produce twice as much content in twice as much time' \cite{orteil}. 


\subsection{Business}
Similar but distinct from economic cost, research is often motivated by higher-level enabling of business models, which we view here as distinct from a technological innovation which changes the specific cost of carrying out a single task or set of tasks. Another distinction is that business-focused research often centres the \textit{player} as a target for optimisation, rather than the \textit{worker}. The description of the IEEE Conference on Games for 2022 describes games as `one of the most profitable industries worldwide'. Papers often cite the huge financial weight of the commercial games industry as a justification for its importance as a target of research \cite{fragoso}.

For example, in \cite{modelling} the authors describe how player retention can be predicted based on the behaviour of players in their first few hours. Such research does not necessarily simplify any existing development task, but might be used in an analysis of the business as a whole, to establish whether a game will become profitable or to enable certain monetisation strategies. Other studies attempt to evaluate the success of business models, for example \cite{cardgamebusinessmodels} evaluates player retention in card games, aiming to impact `revenues and profitability in the \$95billion global games market'. 



\subsection{Capability}
While \textit{Cost} motivations are often concerned with the efficiency of existing processes, many papers also motivate their work by seeking to provide new affordances to AI systems. This might be a new capability for AI but not novel to people, for example in \cite{guzdialvideo} the authors extract game design knowledge from video footage of gameplay, something which humans can do but AI previously could not. They can also be novel capabilities for both AI and people alike, such as \cite{sturtevant} in which an entire generative space is enumerated and used to create intelligent design tools. 

It is generally harder to present work as entirely novel in capability rather than an improvement on existing solutions. The body of published research work is vast, which makes it increasingly likely that a solution has been proposed for a similar problem in a field the author may not be aware of. Novelty is also subjective -- reviewers might not consider existing techniques applied to new domains to be sufficiently novel. At the same time, games are still a young field, and new opportunities for research emerge annually as new genres, platforms and technologies appear, which means novel applications of cutting-edge technology are always available.

\subsection{New Games} \label{sec:newgames}
Research into new technology for games often enables the creation of new genres of game, or a high-concept pitch for a single innovative game idea. This research is less often seen, particularly in AI-focused games conferences such as AIIDE and CoG\footnote{CoG's rebrand has broadened its remit although it is majority AI still.}. We return to the question of why this motivation is less commonly seen at AI conferences later in the paper.

We distinguish \textit{New Games} from \textit{Capability} because capability is often aimed at empowering other whereas game design results in a contribution to games made by the authors themselves. As an example, \cite{innk} presents iNNk, a drawing game in which players collaboratively try to deceive a neural network trained to recognise images.

\subsection{Access}
AI is often applied to partially or wholly automate complex tasks. This results in systems that enable people to engage in activities that might otherwise be closed off to them. In game AI this happens most commonly in two ways: providing wider access to game playing; and providing wider access to game development tools.

In the former case, we see AI techniques being applied to make games more accessible or broaden ways of engaging with games. For example, in \cite{aytemiz} the authors show how an AI player automates certain aspects of game-playing in order for players with different requirements or tastes to engage with the game as best suits them. In the latter case, in \cite{treanorgom} an AI system is aimed at developers with expertise in one area, providing support so they can work in another area.

\subsection{Summary}
The motivations we outline here are, of course, not exhaustive. They are based on our survey of recent papers, textual descriptions of the field by conferences and journals, and our own decade of experience of the field. Broadly speaking, these motivations can be grouped into two categories: expanding \textit{efficiency}, and expanding \textit{capability}. In the former case, we are motivated by a desire to improve some elements of the existing games industry, or the field of artificial intelligence. In the latter, we seek to create new elements, whether that be new types of game or new ways to experience them. In AI parlance, we might consider these approaches to be the complementary ideas of \textit{exploitation} versus \textit{exploration}.

Although we briefly reference it above, note here that we are not distinguishing the target domain. Some papers explicitly mention applications to the games industry \cite{lipsynch} while other papers are interested in applications to broad AI and simply use games as a platform for this \cite{cruz}. We consider work seeking to make a process more efficient to be under the \textit{Cost} category for the purposes of this paper. Later in the paper we examining how external factors drive researchers to do work which benefits different groups.

\section{Motivations for People} \label{sec:personal}
In this section, we step outside of the motivations for individual research projects, and instead examine the motivations of the researchers who conduct the work. While individual research projects or papers are motivated by the specific problems they solve or techniques they propose, the researchers who write them are motivated by complex higher-level motivations that affect their personal lives, career development, and personal long-term research aims and motivations.

As part of this section, we conducted a survey of researcher motivations. We contacted the first authors on every AIIDE 2021 paper and all AI-related COG 2021 papers, and invited them to complete a survey. We also invited them to contact their coauthors to do the same, while anticipating a low completion rate. Of the 114 authors we contacted, 6 emails failed to reach their recipient. We recorded 27 responses in total, a completion rate of 23\%. We include data from this survey throughout this section, although we caution the reader to treat the numbers as anecdotal given the small sample size.

\subsection{Assessing Researcher Motivation}
The sociology of science has long concerned itself with how researchers are motivated and supported, and what consequences this has for their work as a result. In \cite{lam13} Lam identifies three key factors motivating scientific work, which they term \textit{gold}, \textit{ribbons} and \textit{puzzles}. Gold refers to monetary gain of some kind; ribbons refer to prestige, esteem and social standing; and puzzles refer to the curiosity and intellectual satisfaction of doing hard and interesting work.

\subsection{Puzzles}
In the opening part of our survey, we asked participants to state their personal motivations for writing a recent paper, using free text input. This was positioned before questions which asked about specific motivations, to avoid priming the participant. In this first question, around half of participants mention personal interest in the topic as a motivation for the work, with some reporting they were working on research related to games they have played for decades, or published work for no other reason than they were proud of having done it and interested to see what people thought.

Curiosity and interest is part of the popular view of the scientific researcher -- someone driven by a desire to learn about the mysteries of the world around them. There is also a great degree of personal passion found among game AI researchers, which is clear to anyone who has spent a coffee break talking to someone they don't know at a game AI conference. Later in the survey we asked participants to rate how important certain factors were to them in motivating the work they do. For `personal interest', the average score was 4.7/5, the highest-scored factor in the survey.

Despite this, personal interest or curiosity is never mentioned in scientific papers. On the face of it we might find this obvious -- most of us were educated with a certain expectation of how scientific papers are written, and what scientific voice sounds like. It is nevertheless curious to this author that every paper surveyed for this study neglected to mention even a tangential personal interest on the part of the researchers. Personal interest is not simply an emotional sideshow. Having spoken to game AI researchers who are ex-professional gamers, amateur speedrunners, hobbyist developers and lapsed journalists, we know that personal interest brings deep, complex insights that do not fit neatly into other motivational categories. Speaking frankly about what interests us and why is key to revealing new research directions and being honest with one another about what matters to the field.



\subsection{Ribbons}
In his book \textit{Economics: A Very Short Introduction}, economist Partha Dasgupta describes academic research as a system designed around the idea of prestige. In order to encourage scientists to work hard \textit{and} to share the results of their work, academic research has been designed around celebrating the achievements of those who are first to discover, invent and solve. Dasgupta describes this as a `remarkable innovation' because prestige is not costly to provide. By instilling in scientists a desire for this status, this `has enabled [academia] to produce knowledge on the cheap' \cite{dasgupta}. This is not necessarily simply a quest for fame: researchers, especially young ones, need ribbons to survive academia's treacherous career path and secure a permanent job.

Ribbons and prestige do not necessarily mean winning a Nobel prize. Ribbons can come in the form of best paper awards at conferences, invitations to speak at major games industry events, coverage of your work in specialist or mainstream press, or more. A particularly common form of ribbon is to have one's work applied in the games industry. Of the 25 responses to the personal motivation question, five cited dissemination to industry, four cited a desire to see their work applied practically, three cited a wish to do novel work, and three cited benefits to AI as motivations for their work. While some of these motivations are multifaceted (we might want AI to benefit from our work regardless of whether it gains us any prestige) all of these responses can be seen as being motivated by personal impact on the world around us.

Many motivations relating to ribbons and prestige can be seen in papers. As discussed earlier, many authors motivate papers by stating their applicability to either industry or academic research, explaining how they will advance one or the other. Novelty is often a key component of this, for the reasons identified by Dasgupta. There is evidence, however, that a drive for novelty is harming research. In \cite{novelty}, for example, a study of research papers in the field of psychology shows that over time research trended towards more papers claiming novelty, yet these papers also became more cautious and narrower in scope. The desire to have claims of being first to do or solve something leads to a narrower and narrower definition of what novelty is, which harms scientific progress in the long term. A desire to be the first, or best, or most famous, can also encourage plagiarism, exaggerated results, and the prioritisation of speed over rigour. It can also lead to the exploitation of junior researchers, whose contributions are downplayed in order to bolster the profile of senior ones.


\subsection{Gold} 
The games industry, the AI sector, and higher education are all extremely profitable spaces. While money is rarely cited as a direct motivation for research work, performing research can be a means to financial reward, and this is especially true in science and engineering disciplines. Researchers can patent inventions, spin off companies to sell proprietary technology and licenses, or work as consultants on short-term, high-paying jobs for private companies. 

Commercial incentives can greatly impact the nature and direction of research work. It might lead us to shift our priorities, for example refocusing player modelling research on monetisation, rather than player experience. It might lead us to change the ambition of our research, for example by reducing the scope of a content generation system to make it more robust and easier to turn into or embed within a product. In \cite{caulfield} the authors report on a number of studies in the medical field suggesting that governmental emphasis on commercialisation of research negatively impacts research direction, degrades public trust, leads to premature applications of research, and encourages hype in public discussions of science. 


In our survey, participants were asked to score the importance of commercialisation in motivating their work (again scored between 1 and 5). This factor scored the lowest in our survey: 1.93/5 on average, dropping to 1.5/5 for respondents reporting their age as under 30 years old. In their paper on motivation, Lam notes regarding the motivation of `Gold' that while some researchers are proudly motivated by it (citing, for example, `low' university salaries) there was a strong social desirability bias and many researchers did not want to talk about being financially motivated. It is quite possible such a bias exists in our survey as well, although no participant mentioned commercialisation in the free text question preceding the scored category questions. 

There are other reasons to believe that interest in commercialisation is, indeed, low. The games industry, for example, is averse to both risk and workflow disruption, which might make it harder to repackage research for the kinds of company that are usually in talks with universities \cite{pillars}. AI research is also a young field -- PhD intakes have increased drastically in recent years, while senior staff leave for industry. Younger researchers are less able to access commercialisation opportunities due to having fewer connections and more pressure from their career.

\section{Institutional Motivation}
In this section we look at a higher level -- the influences brought by external institutions and organisations, that in turn affect us as individuals, and thus our work. The factors described here are drawn from an analysis of national science policy, funding agency remits, and corporate research goals. In addition to this, we surveyed all grants awarded in the UK by the EPSRC or Innovate UK above £100k in value since 2004, identifying 17 grants in total. While this is an undercount (due to the difficulty of classifying grants by topic, and the limitations of the database), it acts as a useful sample of game AI work being done in a major game AI hub.

Some of these factors will not affect certain researchers -- for example, researchers not working at a university are less affected by their influence. Higher-level motivations in game AI are a Gordian knot of tangled interrelated issues, which means there is a lot of overlap between the areas outlined below. We have roughly broken the topics down by stakeholder, but one of the problems researchers face is the entanglement between the various groups that influence us.

\subsection{Governments}
Governments have two key aims that affect game AI research: first, the growth and sustainability of national games industries; and second the securing of AI research capability. In the former case, governments have taken many approaches to bolstering their games industry -- in Canada, for example, a program of tax breaks and arts funding helped grow the national games industry \cite{gamasutra}. A complementary approach is to provide tax breaks specifically for research and development, or allow access to research funding for games companies engaging in novel work. Access to funding in this way motivates researchers to seek out more immediately-applicable research, and to adjust their working practices to benefit a specific partner. This ties industry applicability to funding, two important influences that can affect career progression.


Governments are also keen to establish a leading role in AI technology. For example, the US National Security Council report on AI demands the US `defend, compete and win in the AI era' \cite{nscai}, while other regions attempt to carve out niches, like the EU's aim to turn Europe into `the global hub for trustworthy AI' \cite{ercai}. Strategic goals such as this inform very high-level funding decisions and long-term strategy. This can motivate researchers to frame their grant proposals in the context of nationalistic competition. In our survey of UK grants, proposals promised to strengthen UK culture, compete with other national research efforts in a topic area, and 80\% of the proposals cite their importance to strengthening the UK economy specifically. Multiple grants list UK government departments or other state institutes as partners, including the Ministry of Defence, demonstrating how game AI interests can easily become aligned with government strategy -- an effect we have seen replicated in many other nations \cite{cook21}.

\subsection{Universities}
\subsubsection{Metrics} Education is heavily metricised in many countries, and these metrics often become fundamental driving factors affecting every decision the university makes. In the UK, metrics for ranking universities greatly impact how students decide where to study, and therefore the university seeks to maximise these metrics at all costs. Currently there are three key rankings: REF, which measures the quality\footnote{We use ``quality'' in the loosest sense here, only because this is the term the UKRI use to describe the aims of their frameworks.} of research; TEF, which measures the quality of teaching; and KEF, which measures `knowledge exchange' which includes commercial collaborations and public engagement.

National ranking systems like REF, along with everyday hiring and funding processes, often explicitly or implicitly rely on secondary metrics such as the CORE rankings and journal impact factors. CORE is a system that assigns a letter grade to every conference, based on a panel. The exact way rankings are assigned is unclear, but their website states it involves a `holistic overview of the data'. In one example on their website, a game AI conference is critiqued because the Program Committee do not have a high enough h-index. 

At the time of writing, no game AI venue has a ranking higher than B in CORE\footnote{CHI Play is an exception -- papers published at CHI Play are actually published in the PACMHCI journal, and so officially has no letter grade.}. Many `top' universities require highly-ranked publications from their staff, which means researchers will either have to publish elsewhere (which will affect the kind of research they can publish, outside of a game AI venue) or may be locked out of jobs at certain institutions, or promotions. This not only affects the individual, but also hampers the influence game AI researchers have as a community in general AI spaces.

\subsubsection{Funding} In many countries universities take a percentage of grant income as `overhead', a figure which varies wildly between different institutions and can be as high as 50\%. As a result, grant-writing is incentivised at universities as it not only allows the hiring of more staff, but also directly provides income to the university. Grant value effectively becomes another metric, as a simple proxy for the value represented by a researcher, and many senior researchers list the value of grant money they have earned on CVs and personal profiles. This amplifies any factors that influence grant funding, including those mentioned elsewhere in this section.

Government funding of public research is rarely adequate, and thus funding must also come from elsewhere. One source can be gifts from large tech firms. Google DeepMind fund several PhD and Masters studentships across the world, for instance. Major firms also act as self-appointed funding agencies, such as Meta's Research Awards scheme. Abdalla and Abdalla, in their study of the influence of Big Tech on AI researchers, found that 84\% of tenure-track computer science faculty at four top universities in the US had been funded by Big Tech at some point in their career, and over half were currently \cite{abdalla}. They show how this influence has been used to change research priorities, identify researchers who are `receptive to industry positions', and to `[groom] academic standard-bearers' who become corporate advocates to students. 




\subsection{Conferences}
As we have already discussed earlier in this section, performance evaluations for researchers often rely on flawed metrics. Many of these metrics are secondary (or even tertiary) measurements based on other metrics, which compounds these issues. These metrics are also motivators for other organisations, such as conferences. Conferences are judged by metrics such as acceptance rate, the citation rate of published papers, and the research profile of those who organise, publish at and attend the conference. Organisers who wish to improve the standing of their conferences may look to optimise one or more of these metrics (a common approach is to reduce acceptance rate by accepting fewer papers, encouraging more submissions, or both). 

Conferences have many ways of shaping researcher motivations. They might introduce special themes which researchers are invited to incorporate into their work (CoG 2022's call was for a focus on AI, while AIIDE 2022 has a special theme inviting papers about negative results). Conferences also define tracks, special sessions and co-ordinate workshops, all of which direct the type of researchers and research that appears at the conference.

Conferences also implicitly affect the behaviour of researchers through the way they are structured and organised. During the first two years of the COVID-19 pandemic, many events switched to remote participation only, which provided greater access to those who would not ordinarily be able to travel. The long-term effects of this are yet to be seen, but if events switch back to fully in-person (as some seem to be edging towards) this will accordingly affect who can publish at them. This can encourage researchers to prioritise journal publications, for example, which do not require travel visas or funding to publish work with. 

\section{A Personal Case Study}
In this section I will provide a brief reflection on my own motivations as a researcher, and how and why they have changed over the last decade. I provide this here partly as a way to explore how the motivations of researchers can be affected by factors which do not come across in papers, but also to help motivate \textit{this} paper in itself by showing how I came to struggle to find justifiable motivations for my research. This section sadly had to be cut from the submitted version of this paper, as the paper was already far beyond its page limit\footnote{Cutting a section about personal background from this paper in order to get it published felt quite poetic.}.

\subsection{Early Work}
ANGELINA is an AI system for automating game development. It was designed to produce games autonomously, but research work on the system has applications to co-creativity as well. Research on the system covered a number of topics in game AI: generative approaches to game design; general game-playing techniques for evaluating games; computational creativity perspectives on creative AI; logic and constraint solving for shaping the design space; and data manipulation and knowledge extraction techniques for enhancing the games themselves.

Initially, my motivation for the work was firmly in the realm of Lam's `puzzle' category -- I was interested in the feasibility of building a system that could generate games, as a challenging problem and a way of exploring creativity. My first paper motivates itself by describing generative approaches in a general sense, noting that they can create `more dynamic and interesting games' (c.f. \textit{New Games}, \S \ref{sec:newgames}) and that generative systems can help `alleviate the burden on developers' and `tackle the content generation problem' (c.f. \textit{Cost}, \S\ref{sec:cost}) \cite{angelina1}. It is worth noting that I provided no evidence for the claim of alleviating developer burden. As discussed earlier, this claim is often used without justification.

I also note that my initial motivations partly came from a place of uncertainty. As a junior researcher, entering a new community (and career) it was unclear what the boundaries of acceptability were. I found it safest to adopt motivations I perceived as having strong foundations in existing work in the area -- thus, I cited applicability to the games industry and their business practices. Nowhere is there mention that I simply thought this was an interesting line of research to pursue, or that it might lead to novel future research questions. This was the beginning of a process of hiding or reshaping motivations for specific academic audiences, a practice I know is well-known to many reading this, as most academics discuss this as a simple fact of publishing research or writing grant applications.

\subsection{Developing Motivations}
Over the course of the PhD I became more comfortable both with conducting research and with communicating that research with my peers. I visited many conferences (I was fortunate enough to be well-funded and supported by my supervisor) and gained an understanding of where and how community norms could be bent. For example, while my papers retained a more formal voice, my talks because increasingly informal and relaxed. Often I would give additional motivations in my talks that did not appear in my papers (which creates a problem for the long-term record of the field, as talk recordings will likely not survive, and many were not recorded at all).

As I became more comfortable with honestly sharing my motivations and opinions, those motivations also shifted and changed. My work had broadened to include more subjective problems: conveying meaning through game design; understanding and incorporating art and sound; and engaging with communities of designers. We can think of these problems as new kinds of `puzzle'. Papers I published in this period began to drop their motivations altogether. In \cite{ludusexmachina} for example, I simply observe that game development is a challenging creative task for AI researchers to tackle, and then explain how we have taken a step towards doing so. This is partly enabled by writing `system description' papers which allow for a system to be presented with less emphasis on evaluation -- which, in turn, de-emphasises solving a specific problem and allows us to speak more generally about engineering a system. 

Finally, another key change in this period was a need to think about my career. Every academic career path is different, and many academics will give unique (and often conflicting) advice on how best to get a job in research. The need to develop a profile that would let me apply for funding or a permanent job would have affected many of the decisions I made during this time, consciously and unconsciously. For example, some papers I published during this time were written not because I felt they \textit{needed} to be written, but because I knew they would strengthen my case as a researcher applying for a job. Recall earlier that 75\% of under-35s in research are on temporary contracts -- many of these researchers are under similar pressures. I was also forced to make difficult decisions due to family commitments, which increased the pressure to find employment in specific locations or with remote work.

\subsection{Motivations Today}
Today I have survived through the most stressful parts of the academic career ladder\footnote{Entering this into the public record of the field intentionally so we can all laugh at it in twenty years' time.}. Although publish-or-perish pressures still exist, my job is more secure than ever, and I feel comfortable within the game AI community (comfortable enough to write a strange meta-paper about researcher motivation). This security also allowed me an opportunity to rethink the direction of my research, and pursue research leads that I might have previously thought did not have a sufficiently strong motivation, but am personally interested in \cite{cook19}.

Of course, as old pressures disappear, new ones emerge. My time is now more constrained due to the responsibilities of a university posting, which forces prioritisation and thus amplifies the pressure applied by externals such as funding agencies or my employer. I also supervise a group of brilliant students, and I have a duty to ensure they are able to achieve their own goals, whether that is to gain employment after their PhD, take up a career in academia themselves, or something else entirely. This might override my own personal motivations at times -- for example, while I might decide I do not care about high-ranking journals, this decision may be harmful if I apply it to my student's work, too.

In attempting to write the first paper about my new system, Puck, I realised that it was harder than ever to justify or explain my motivations for the work. In some cases, I was willing to explicitly work against traditional motivations (like an increase in efficiency) in order to achieve other goals. In the end, I decided not to submit the paper to the IEEE Conference on Games to give me more time to rewrite it. The experience of thinking about how to justify the work, and realising how difficult it was becoming, led me to reflect on how we justify our work, and thus to this paper you are reading now.

\subsection{Reflexive Statement}
Following a recommendation by Dr. Jamie Woodcock, below is a brief \textit{reflexive statement}, a technique used in sociology and related fields to provide background on the author. Hopefully this will provide context to the perspective I have brought to this paper, especially if your perspective on or experience of academia differs greatly from mine.

Although I had no experience of computing prior to university, I studied at a well-funded and prominent university, giving me privileged access to good PhD opportunities, and spend the first half of my career so far supported by a well-funded supervisor. Family commitments restricted where I could apply for academic jobs, and led me to live abroad for several years -- without my privileged start in academia, it would likely not have been possible for me to stay in a research job. Some members of the community will be more privileged still, and some less, although there is a degree of survival bias in academia which makes it hard for those without a baseline amount of privilege to easily remain in the sector without enduring enormous pressure.

I am currently funded by the Royal Academy of Engineering. This provides me a lot of independent funding and support, and makes me much more secure than other researchers on fixed-term contracts. The RAE themselves are a complex organisation -- they have strong advocacy and support programs for minority groups and the Global South; however, they also partner with arms manufacturers and the `UK Intelligence Community'. Although these partnerships are unrelated to my funding, this shows the complex web of partnerships that permeates academic funding bodies, and I do not wish to present myself as someone immune to this.

I am a white man, who grew up speaking English as a first language, and thus have had a mostly privileged path in my career so far. I have become increasingly left-wing in my political outlook over the last fifteen years, and generally advocate against the dominance of large corporations in the AI sector, as well as arguing a number of other issues that would be seen as highly political. All of these facts had a clear impact on my decision to write this paper, and the issues I chose to frame within it. I have tried to carefully provide evidence for claims made here, partly to mitigate the sense of bias that my personal views might bring.

I wrote this paper partly as a personal exercise to both reflect on my own views on scientific research, and to collect together perspectives and studies surrounding this issue. However, this paper is also part of a wider belief of mine that our field (and indeed, STEM research as a whole) can and should upend many of its core beliefs and practices, in order to build a more just, more effective and more ambitious research community that better serves the public who fund it.

\section{Implications}
Throughout the paper so far we have tried to illustrate the wide variety of factors that influence the nature of game AI research done today, from personal interests through to international strategic co-ordination and market influence. In this section we now ask what the consequences of the current status quo are, and what issues might arise from them.

\subsection{Undersupported Research Directions}
Throughout this paper we have shown that game AI researchers are often influenced in their work by factors other than their own personal sense of what is the best next step for research. It stands to reason, therefore, that some areas that would be beneficial to game AI research are receiving less attention than they deserve.

For example, a recurring theme throughout this paper is how various influencing factors lead to a large number of collaborations and partnerships with large commercial game developers. Funding agencies and national strategy encourages industrial applicability, but this applicability favours larger commercial game developers due to their financial capacity to engage in R\&D and the increased weight of having an important company as a partner. This raises the question: what research would be of benefit to solo game developers with little or no budget? What research would be of benefit to game developers who have no interest in economic impact? These research questions are harder to justify answering than those which directly impact AAA game development.

Earlier we discussed how a reliance on metrics for evaluating researchers, combined with the precarity of modern research careers, has led to more cautious attitudes from researchers. For example, Super Mario is often chosen as a domain for level generation research\footnote{This is not a criticism of research focusing on Super Mario, and other factors are at play here (such as a need to compare work). Nevertheless, we believe the factors described here influence our decisions too.}. Is this the case because it is the best way forward for level generation research or, at least partly, a consequence of huge pressure on researchers to deliver reliable, publishable, conclusive research while limited on time and resources? It is up to us as a community to ask these questions and make adjustments if we believe necessary.


\subsection{Overemphasised Research Directions}
As a complement to the previous section, strong external influences can warp some research areas and result in too much focus. In the most simplistic sense, this is evident when researchers collaborate with a single company on a very specific problem that does not offer much benefit to the field as a whole, which can lead to researchers acting more akin to private consultants than public researchers. We can also see this manifest in research trends, where a technique explodes in popularity and interest. The AI boom may have led to what Klinger et al. describe as a `narrowing of AI research' in recent years, as researchers conservatively focus on popular techniques and domains, at the expense of exploring alternatives \cite{klinger20}. This narrowing, coupled with the ubiquity of a few machine learning platforms and techniques, has parallels to the standardisation of game development through middleware, which had its own narrowing effects on creativity \cite{woodcock}.

As an example of this, consider machine learning in procedural generation. Clearly, machine learning has grown in popularity over the last decade, and is possibly one of the most incentivised research topics in research, not just in computer science. In 2012, no procedural generation paper published in IEEE CIG used machine learning. In 2016, 2 of the 8 papers on the topic use machine learning. In 2021, 12 of the 18 papers used machine learning techniques, with another 2 we consider borderline. The novelty and exciting results from machine learning undoubtedly drive some of this interest and growth, but we must also accept that part of its proliferation is down to the many factors we have discussed in this paper, for it to comprise two thirds of a field's published research.


To be clear: this is \textit{not} about mandating who can or cannot research which topics. This is about reflecting on which topics we are encouraging and incentivising the study of, and the risk of imbalancing our priorities as a community without realising. It might be that we conclude there are no such issues. But given the vast web of external influences, that seems unlikely.

\subsection{Dominant Influence of Capital}
While game AI research contains many diverse and unusual research projects, it is clear from the examples we have outlined throughout this paper that there is a dominating focus on commercial industrial applications and economic benefits, as well as a disproportionate influence brought by large, capital-driven private research entities.  Governments, funding agencies and research partners emphasise productivity, efficiency, cost and scale. A cursory survey of papers and grant applications in our field by an outsider would give the impression of a field obsessed with economics.

There is nothing inherently bad about improving the efficiency of an algorithm or reducing the cost of a task, and these remain useful motivations for scientists to have. What complicates matters is when we research in an environment dominated by large corporate concerns, be that FAANG-style Big Tech firms or blockbuster game developers. In this context, our research is more likely to enable the worse elements in these spheres, help entrench and focus power, and specifically optimise the efficiency of existing capitalist entities. We should be wary of the \textit{kinds} of efficiency we are enabling, and how that contrasts with our intent. When we motivate automation by saying it removes boring work, how sure are we that this is the case? We must do better to reflect on these issues in our work that are glossed over by repeating common motivations.

\subsection{Researcher as Practitioner}
Games lie at an awkward intersection between art and industry. Attending conferences about AI and music, which is more comfortably recognised as an artform, it is more likely to see speakers discuss their own personal musical practice, how it influences their research, and vice versa. Researchers in the visual arts are often recognised as artists in their own right, and it is understood that they are able to make contributions to the world directly through their research. Game AI developers have won awards at prestigious games industry events for games they have made \cite{badnews} \cite{promweek}, yet in general there is not a culture of celebrating or encouraging this within the field.

The emphasis on applications to industry, support of large commercial entities, and economic impact has led to an established trend of the game AI researcher as a subordinate to the games industry. This is not mandated or enforced anywhere, yet we also do little to encourage otherwise. This year we are pleased to see AIIDE introduce a new track which allows full proceedings publications for work describing the development of a game which incorporates novel AI techniques. We know from experience that game AI researchers can build innovative games, that this is a major contribution to the field, and it can reveal new research too. Yet our field, particularly the harder AI side, is not designed to encourage it.

\section{What Can Be Done?}
\subsection{Collective Action}
In a traditional industrial setting, workers wishing to change something about their circumstances might collectively take action to bring about that change. A complication faced by researchers working in a field together is that they are widely distributed across the globe, with only a handful of them sharing the same institution or even country. While many researchers are unionised in countries around the world, the role of those unions is typically to campaign for issues affecting the university or education sector, rather than to tackle specific issues relating to one research field.

The state of powerlessness that researchers find themselves in is, in part, due to the structure of academia. As described in section \ref{sec:personal}, academic research is built on a principle of competition, meaning we are to some degree incentivised to avoid collective action that might harm our individual success, even if it means collective benefit. Similarly, our being globally distributed makes it hard to co-ordinate action against specific issues within our field. For example, if AI researchers at my institution were to reject funding from Meta or DeepMind, those companies would simply look to other institutions for collaborators. Only by co-ordinating action together would we be able to push back against some of the external forces affecting our field (should we wish to do so).

Collective action is also important to address the inequality and prejudice marginalised researchers experience. Systemic bias and oppression can warp and amplify the negative effects of some of the factors explored in this paper (for example, the gender citation gap amplifies the negative impact of publication metrics \cite{women}). An equitable and just scientific community is a good in and of itself, of course, but it is also fundamental to empowering our community to take control of its future.

We realise that the word `union' is, in some circles, a controversial concept whose meaning has been twisted by decades of negative news and politics. Nevertheless, we must face the fact that it will be hard to change our field for the better without collective action of some kind. Researchers are not strangers to banding together to build things -- our conferences, our journals, most of our infrastructure is built and maintained by our own communities. It is a natural extension of this communal spirit to act together to rebalance our field's incentives and goals, too.

\subsection{Encouraging Honesty}
One of the conclusions we drew from researching and writing this paper is that incidental motivations help contextualise work in a way that `official' motivations do not. It is evident from countless conversations with researchers, as well as the anecdotal support provided by the surveys we conducted, that academics frequently motivate their work using reasoning that they themselves are not actually motivated by, or at the very least omit motivations they are concerned are not acceptable. This is because they understand that political games must be played in order to get hired, funded and published. We are taught about these games by our supervisors, our mentors, and our colleagues as we progress through our career.

Obscuring our motivations is not necessarily harmful, but it can reduce our effectiveness as a scientific community, in two ways in particular. Firstly, if we pretend our work is motivated by certain factors, we normalise and elevate them. This influences the perceptions of new community members, signals to outsiders what our field prioritises, and may affect decision-making by key stakeholder groups (for example, more senior researchers might encourage research in a particular direction simply because it seems to be the prevailing trend in a field). Secondly, it obscures potential leads and insights that are not captured by acceptable motivations. Personal conversations with researchers reveal colourful motivations and excitement about their projects, yet these so often do not make their way into the permanent record of the field. 

It isn't possible for us to suddenly stop this practice altogether -- some political games must still be played. But perhaps at least in venues that we wholly own and organise, such as IEEE CoG, we can encourage a more open discussion of what we find valuable and why.  We do not know how best to encourage this, but perhaps this paper in itself might help a little in that regard. This is, first and foremost, a cultural problem, and hopefully by discussing and making small changes we can begin to adjust this.

\subsection{Improving Our Community}
Many of the personal and institutional effects identified in this paper do not affect us all equally, and disadvantaged groups experience negative effects more harshly. For example, while all of us are caught up in the obsessive measurement of citations and impact factors, minority groups are affected much more strongly due to effects such as the citation gap \cite{race}. Being able to take risks, pursue research of our choice, or turn down unwanted partnerships is easier with privilege, financial security and social standing.

One way to help give our community back a sense of independence and control over their work and career is to help address the fundamental problems relating to equality and fairness \cite{cook21}. Fighting prejudice and bias, helping combat systemic inequality in academic spaces, rooting out abusers from our communities -- this helps empower and strengthen all of us, not just those directly affected. Empowered scientific communities are better able to decide their own fate, and this ties directly to our ability to push and pull against or with the various influences we have described in this paper. Stronger communities get to dictate more of their own future.

\subsection{Broadening Our Network}
Access to resources is a common thread running through many of the external influences in this paper. Computer science research is relatively inexpensive to conduct (at least, prior to the age of the GPU cluster) but still requires significant funds to employ people and support them in engaging in activities like publishing and presenting work. Finding new ways to fund this work can allow us to escape certain external pressures (the trend of accepting funding from Big Tech is itself a way to escape the whims of funding agencies, albeit at great cost).

In order to do this, we must engage with new communities, new stakeholder groups, and establish new support structures. We believe that the most immediate way we can do this is through our conferences and seminars, by co-locating with new communities (as far as we can tell, IEEE CoG has not been co-located with a conference in the last decade) and inviting participation from new domains (CoG's keynotes are largely either AI researchers or AAA game developers, for instance). Organising meetings with new communities outside of academia would also be helpful -- learning about how other groups fund and support themselves and their communities will reveal new ways we can seek support.

\section{Conclusions}
\begin{quote}
\textit{The popular stereotype of the researcher is that of a skeptic and a pessimist. Nothing could be further from the truth! Scientists must be optimists at heart, in order to block out the incessant chorus of those who say ``It cannot be done.''} \cite{alphacentauri}
\end{quote}

In this paper we have discussed the motivations and influences that affect game AI research. We have argued that current influences drive us towards a particular type of research, with an excessive influence from large companies, and overly focused on taking a supportive, rather than leading, role.

We are not aiming to change anyone's motivations by writing this paper. However, researchers (and research) are subject to many complicated influences, many of which are not often discussed. The AI boom and marketisation of higher education have also drastically changed some of these factors in the past decade, and a lot of our communal institutions have not adapted to this new landscape. Millions of dollars of research money and decades of person hours are being spent on game AI research every year, and it is therefore vital to think and talk about what is driving the direction of our field, and what we are aiming for.

In 1942, the sociologist Robert Merton identified what he saw as four critical features that scientific communities must aspire to, now known as \textit{Mertonian norms} \cite{merton42}:

\begin{itemize}
\item \textbf{Communism} -- the notion that scientists share ownership of research, collectively.
\item \textbf{Universalism} -- the notion that the traits of a researcher do not affect how the research is judged.
\item \textbf{Disinterestedness} -- the notion that scientists should work for the benefit of science, rather than personal gain.
\item \textbf{Organised Skepticism} -- the notion that scientific claims must be questioned on an institutional level. This is often interpreted as being a call for peer review and critical evaluation by the community.
\end{itemize}

Merton's norms are not perfect by any means, and certainly a product of the time in which they were written, as scientists re-examined their role in the world during World War 2. Yet we can see many of these issues crop up today, in both good and bad ways. The push for open access research and the sharing of data is emblematic of Merton's Communism, yet the increasing weight of private research labs and secrecy erode the fragile communal spirit already in place. It is both heartening and depressing to see that the issues we face today are longstanding. It means they are hard to beat, but it also means that scientists have always strived to overcome them and build better communities, and we can too.





\section{Acknowledgements}
Thanks to Julian Hough, Michaela MacDonald and everyone in the QMUCU for their support during this year. Thanks to Georg von Graevenitz for help understanding the UKRI grants system. Thanks to Stuart Watson, Sam Geen and Kor Bosch for good conversations about academia's bizarre excesses, and to Anndra Dunn for encouraging me to quote Alpha Centauri. Many thanks to Azalea Raad, Florence Smith Nicholls and Jamie Woodcock for providing feedback on an early draft of this paper -- thanks to Jamie also for continuing to be a great teacher and friend. Thanks to the discussion groups who hosted me to discuss the 2021 Social Responsibility paper, in which I was also able to raise some of the issues here and get new perspectives on them. And thanks to the Knives \& Paintbrushes team, Mads, Youn\`{e}s and Florence, who are always inspiring.

\bibliographystyle{plain}
\bibliography{cog22}

\end{document}